\documentclass[sort&compress,
    ,final            
  ]
  {aipproc}
  
\usepackage{amssymb}
\RequirePackage{xspace}
\def\Vub  {\ensuremath{|V_{ub}|}\xspace}
\def\Vcb  {\ensuremath{|V_{cb}|}\xspace}
\def\btoulnu   {\ensuremath{\lowercase{b} \to \lowercase{u}\ell \bar\nu}\xspace}
\def\btoclnu   {\ensuremath{\lowercase{b} \to \lowercase{c}\ell \bar\nu}\xspace}

\def\calrmB{\hbox\bgroup B\egroup}

\newboolean{beauty}
\setboolean{beauty}{true}

\layoutstyle{8x11double}

\begin{document}


\title{\hfill The status of $\Vub$\footnote{
The following article has been submitted to the Proceedings of Beauty 2003.
After it is published, it will be found at
\protect\url{http://proceedings.aip.org}.}\hfill\makebox[0pt]{\raisebox{5ex}[1ex]{{\normalsize CLNS 04/1863}}}}

\author{Lawrence Gibbons 
\\ {\normalsize\sl (Representing the CLEO Collaboration)}
}{
address={Cornell University, Department of Physics, 
         Ithaca, NY 14850, U.S.A.}
}

%

\begin{abstract}
I survey the theoretical and experimental information available for determination
of $\Vub$ with inclusive and exclusive techniques.   Using recent
experimental and theoretical advances,  I outline a  procedure in which the
inclusive information can be combined to obtain an inclusive $\Vub$ that includes 
experimentally--derived uncertainty estimates for outstanding theoretical corrections.

\end{abstract}

\maketitle


\subsection{Introduction}

The magnitude of the Cabibbo--Kobayashi--Maskawa 
(CKM)\citep{Cabibbo:1963yz,Kobayashi:1973fv}  matrix element $V_{ub}$ remains 
a crucial input into tests of the unitarity of the 3-generation CKM matrix,
yet a $\Vub$ averaging procedure that results in a defensible 
uncertainty remains elusive.   Recent experimental 
and theoretical progress provides us with an opportunity to improve this situation.  
Here I review the theoretical and experimental concerns in extraction of
$\Vub$, and outline a potential inclusive ``averaging'' procedure 
in which information from all regions of phase space can be combined to bound
experimentally the missing theoretical contributions that plague
our averaging attempts.

\begin{figure}[h]
 \includegraphics[height=.20\textheight]{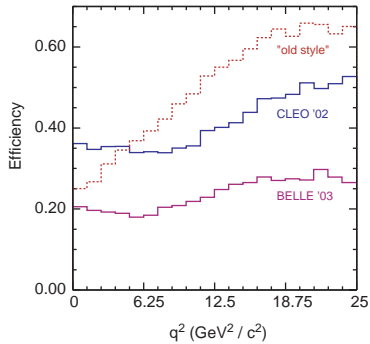}
  \caption{Efficiency as a function $q^2$ for the ``traditional" lepton endpoint analysis
  (dotted curve) and more recent CLEO and BELLE analyses (solid curves).}
  \label{fig:endpoint_eff}
\end{figure}

Inclusive theory has progressed
significantly in categorization of the important corrections in different
regions of phase space, and in determination of some of the hitherto unknown corrections.
Experimentally, measurements have significantly reduced model
dependence in extraction of rates.  For example,  the reduced dependence of  efficiency  on
dilepton mass $q^2=M_{W^*}^2$  in new rate measurements near the $E_\ell$ 
endpoint\citep{Bornheim:2002du,Aubert:2002pi,Abe:2003endpoint} 
(Figure~\ref{fig:endpoint_eff}) has reduced model dependence by a factor of three.
Recent measurements
have also minimized reliance on detailed $d\Gamma(B\to X_u\ell\nu)/dE_\ell dM_{X_u}dq^2$
modeling to separate the $b\to u\ell\nu$ signal from
$b\to c\ell\nu$.  We thus have rate measurements with well-defined sensitivities
within well-defined regions of 
phase space 
for which we can categorize the important theoretical uncertainties.   
Both the theoretical and experimental improvements are key to more
robust evaluation of $\Vub$.

\subsection{Inclusive  $\btoulnu$}

Theoretically, issues regarding the calculation of the total semileptonic 
partial width $\Gamma(B\to X_u\ell\nu)$ via the operator product expansion (OPE)
are  well-controlled\citep{Bigi:1993ex,Neubert:1996qg,Bigi:1997fj,Hoang:1998ng,Bigi:1999dv,Ligeti:1999yc}. 
This nonperturbative power series in $1/m_b$ and perturbative
expansion in $\alpha_s$ has,
  at order $1/m_b^2$, two nonperturbative parameters: $\lambda_1$ or $\mu_\pi^2$,
which is related to the Fermi momentum of the $b$ quark in the meson, and $\lambda_2$ or
$\mu_G^2$, which parameterizes the hyperfine interaction between the heavy
 quark and the light degrees of freedom.  The $\lambda$ and $\mu$ parameters differ in their
infrared behavior.  In terms of these parameters, the OPE at $1/m_B^2$ yields
\begin{eqnarray}
\label{eq:parton}
\lefteqn{\Gamma(B\to X_u e\nu) = {G_F^2 |V_{ub}|^2\over 192\pi^3}\, m_b^5} \\ \nonumber
& & \times\left[1-
\frac{9\lambda_2 - \lambda_1}{2m_b^2} +\ldots - {\cal O}(\frac{\alpha_s}{\pi})\right].
 \end{eqnarray}
The perturbative corrections are known to order $\alpha_s^2$ \citep{vanRitbergen:1999gs}.
The error induced by uncertainties in the nonperturbative
 parameters $\lambda_{1,2}$ is relative small, and an evaluation by the LEP Heavy
 Flavors \citep{hfworkinggroup} working group yielded
 \begin{eqnarray}
 \label{eq:rate}
\lefteqn{\Vub = 0.00445
\left( 
\frac{{\calrmB}(b\to u\ell\bar{\nu})}{0.002} \frac{1.55\mathrm{ps}}{\tau_b}
\right)^{1/2}} \\
& & \times \left(1\pm 0.020_{\lambda} \pm0.052_{m_b}\right) \nonumber
\end{eqnarray}
The mass $m^{1S}_b (1GeV ) = 4.58\pm0.09$ GeV, which agrees with
a recent survey\citep{El-Khadra:2002wp}, dominates the uncertainty.
Because the OPE is a quark-level calculation, 
the estimate rests on the assumption of global quark-hadron duality,
which  is well-motivated\citep{Bigi:2001ys} 
for  $\btoulnu$ (particularly with its broad range of hadronic
final states).   We can bound $\sigma_{\mathrm{duality}}$ by comparison 
of the more precise  inclusive and exclusive evaluations of $\Vcb$. 
The difference\citep{Artuso:2004rpp} of $(0.8\pm 1.6)\times 10^{-3}$ implies
$\sigma_{\mathrm{duality}}=4\%$, for a total uncertainty of 6.8\% on
the total rate.

\begin{table}[b]
\ifthenelse{\boolean{beauty}}{
\begin{tabular}{crrl} \hline
$p_\ell^{\mathrm{min}}$ &	$\Delta {\cal B}_u(p)$ &	\multicolumn{1}{c}{$f_E$} \\ 
(GeV/$c$) & ($10^{-4}$) & & \\
\hline
 2.0	& $4.22 \pm 1.81$   & $\dagger 0.266 \pm 0.048$ & CLEO ($e,\mu$) \\
 2.1	& $3.28 \pm 0.77$   & $\dagger 0.198 \pm 0.040$   & CLEO ($e,\mu$)\\
 2.2	&  $2.30 \pm 0.38$   & $\dagger 0.130 \pm 0.028$   & CLEO ($e,\mu$)\\
 2.3	& $1.43 \pm 0.16$   & $\dagger 0.074 \pm 0.017$   & CLEO ($e,\mu$)\\
    	&  $\dagger 1.52 \pm 0.20$   & $ 0.078 \pm 0.017$   & BaBar ($e$)\\
  	&  $1.19 \pm 0.15 $   & $\dagger 0.072 \pm 0.016$   & BELLE ($e$)\\
 2.4	& $0.64 \pm 0.09$   & $\dagger 0.037 \pm 0.008$   & CLEO  ($e,\mu$)\\
\end{tabular}
}{
\begin{tabular}{crrl} \hline
$p_\ell^{\mathrm{min}}$ &	$\Delta {\cal B}_u(p)$ &	\multicolumn{1}{c}{$f_E$} \\ 
(GeV/$c$) & ($10^{-4}$) & & \\
\hline
 2.0	& $4.22 \pm 0.33 \pm 1.78$   & $\dagger 0.266 \pm 0.041 \pm 0.024$   & CLEO ($e,\mu$) \\
 2.1	& $3.28 \pm 0.23 \pm 0.73$   & $\dagger 0.198 \pm 0.035 \pm 0.020$   & CLEO ($e,\mu$)\\
 2.2	&  $2.30 \pm 0.15 \pm 0.35$   & $\dagger 0.130 \pm 0.024 \pm 0.015$   & CLEO ($e,\mu$)\\
 2.3	& $1.43 \pm 0.10 \pm 0.13$   & $0.074 \pm 0.014 \pm 0.009$   & CLEO ($e,\mu$)\\
    		&  $\dagger 1.52 \pm 0.14 \pm 0.14$   & $ 0.078 \pm 0.015 \pm 0.009$   & BaBar ($e$)\\
  		&  $1.19 \pm 0.11 \pm 0.10$   & $\dagger 0.072 \pm 0.014 \pm 0.008$   & BELLE ($e$)\\
 2.4	& $0.64 \pm 0.07 \pm 0.05$   & $\dagger 0.037 \pm 0.007 \pm 0.003$   & CLEO  ($e,\mu$)\\
\end{tabular}
}
\caption{Partial branching fractions for $\btoulnu$ near the $p_l$ endpoint. The
rates are integrated up to $p_\ell^{\mathrm{max}}=2.6$ GeV/$c$
($\Upsilon(4S)$ frame). The estimated rate fraction $f_E$  is given for each range.  
The dagger ($\dagger$) indicates where the QED radiative correction was applied. }
\label{tab:endpoint_rates}
\end{table}

To overcome
the 100 times larger $\btoclnu$ background, inclusive $\btoulnu$ measurements 
utilize restricted regions of phase
space in which the background is  suppressed.
Extraction of $\Vub$ then requires the fraction of the 
total $\btoulnu$ rate that lies within the given region of phase space. This complicates
the theoretical issues and uncertainty considerably.
We  first consider measurements restricted either to the 
region ${p_\ell}\gtrsim 2.2$ GeV$/c$ or to
the region of low hadronic mass  ($M_X\lesssim M_D$).  Within both, the
parton--level OPE fails because its 
expansion parameter ${E_X\Lambda_{QCD}}/{m_x^2}\sim 1$.  
We will return to this issue and its impact on rate estimations.
 
 Table~\ref{tab:endpoint_rates} summarizes the rate measurements by 
 CLEO\citep{Bornheim:2002du}, BaBar\citep{Aubert:2002pi} 
and Belle\citep{Abe:2003endpoint}  near the $p_\ell$ endpoint.  These analyses must suppress background
from continuum processes, and can reach down to about 2.2 GeV$/c$ before control
of $b\to c\ell\nu$ becomes problematic.  The continuum suppression
induces  the efficiency variation with $q^2$ discussed above. The  remainder of the model dependence could be eliminated by coarsely binning in $q^2$, or, preferably, through
use of a tagged $B$ sample.

\begin{figure}
 \includegraphics[width=0.77\columnwidth]{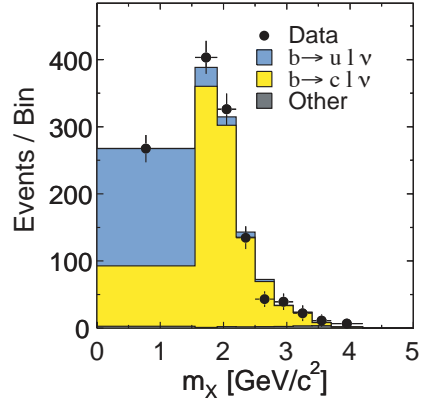}
 \source{\citet{Aubert:2003zw}}
  \caption{Reconstructed $M_X$ distribution from the BaBar low $M_X$ analysis.}
  \label{fig:babar_mx}
\end{figure}

BaBar\citep{Aubert:2003zw} and Belle\citep{Schwanda:2003eps} have
new analyses of the low $M_X$ region (combined with a moderate $p_\ell>1.0$ GeV$/c$
requirement), which was first studied by DELPHI\citep{Abreu:2000mx}.
Finite $M_X$ resolution smears the $\btoclnu$ background
below its theoretical limit of $M_D$, forcing more stringent $M_X$ requirements that 
approach a pole in the parton level spectrum.
The preliminary Belle analysis utilizes a $D^{(*)}\ell\nu$
tag, with $M_X$ calculated directly from all particles after removal of tag and lepton.
This tag still results in a significantly larger background than signal level,
and the systematic estimates (very preliminary) appear quite aggressive.
The BaBar analysis uses
fully reconstructed hadronic $B$ tags, again with direct $M_X$ calculation.
This analysis  achieves a beautiful $\btoulnu$ signal 
in the region $M_X<1.55$ GeV$/c$ with signal to background
ratio (see Figure~\ref{fig:babar_mx})
approaching 2:1. This ratio shows  the anticipated power
of the hadronic $B$ tags, which afford unsurpassed resolution on $M_X$.
The efficiency versus $M_X$, while not featureless, appears promisingly uniform.  
BaBar has also extracted the yields with a fit to the integrated
$M_X<1.55$ GeV$/c$ interval.  Both features minimize
dependence of the extracted rate on detailed modeling of the $\btoulnu$,
which allows  for cleaner theoretical interpretation and simplifies
improved determination of $\Vub$ as theory advances.   

\begin{figure}
 \includegraphics[width=\columnwidth]{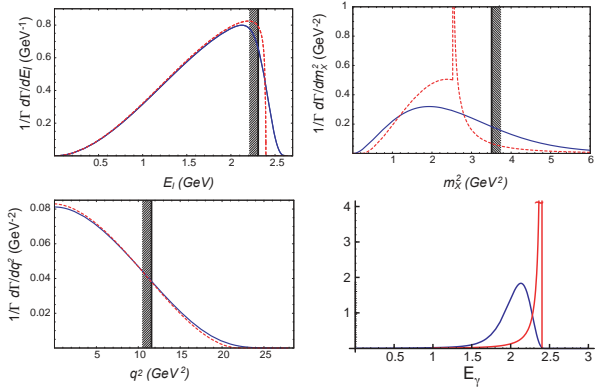}
  \caption{Parton level (blue) and model convoluted (red) spectra for 
  $B\to X_u\ell\nu$ (top, bottom left;
  from ref.~\citep{Bauer:2000zb}) and for $B\to X_s\gamma$ (bottom right).}
  \label{fig:convolved}
\end{figure}

Determination of the fraction of the $\btoulnu$ rate in the $p_\ell$ endpoint or the
 low $M_X$ region requires resummation of the OPE to all orders in 
 ${E_X\Lambda_{QCD}}/{m_x^2}$\citep{Bigi:1993ex,Neubert:1994ch,Neubert:1994um,Dikeman:1996ad,Aglietti:2000te}.
 The resummation results, at leading-twist order, in a nonperturbative shape
 function  $f(k_+)$. The argument
 $k_+ = k^0 + k_\|$, where $k^\mu = p_b^\mu - m_b v^\mu$ is the $b$ quark momentum with the ``mechanical'' portion subtracted. The spatial components $k_\|$ and $k_\perp$ are
 defined relative to  the $m_b v^\mu - q^\mu$ (roughly  the recoiling $u$ quark) direction. 
 At leading twist,  effects like the ``jiggling'' of $k_\perp$ are ignored, and the
 differential partial width is given by the convolution of the shape function with the
 parton level differential distribution (see Figure~\ref{fig:convolved}):
 \begin{equation}
 \label{eq:convolve}
d\Gamma =\int dk_+\,f(k_+){d\Gamma^{\mathrm (parton)}_{m_b\to m_b+k_+}}.
\end{equation}
Because the shape function depends only on parameters of the $B$ meson, this description
holds for any $B$ decay to a light quark, such as $B\to s\gamma$.

It has been known for some time\citep{Bigi:1993ex,Neubert:1994um} that measurement
of the $E_\gamma$ spectrum in $b\to s\gamma$ can yield (at leading twist)  $f(k_+)$.  Ideally, $\Vub$ would be determined directly from integrated 
spectra\citep{Neubert:1994um,Leibovich:1999xf,Leibovich:2000ey,Neubert:2001sk,Leibovich:2001ra,Bigi:2002qq}
without extraction of an intermediate shape function. 
For the lepton spectrum, for example, one takes 
\begin{equation}
\label{eq:integrate }
\left| \frac{V_{ub}}{V_{tb} V_{ts}^*} \right|^2 
   = \frac{3\alpha}{\pi}\,K_{\rm pert}\,
   \frac{\widehat\Gamma_u(E_0)}{\widehat\Gamma_s(E_0)}
   + O(\Lambda_{\rm QCD}/M_B) \,,
\end{equation}
where $K_{\mathrm pert}$ is a calculable perturbative kernel, and $\widehat\Gamma_u(E_0)$ and $\widehat\Gamma_s(E_0)$ are appropriately
weighted integrals over, respectively,  the $E_\ell$ and $E_\gamma$ spectra
above the cutoff $E_0$.  A similar expression exists for the
$M_X$ spectrum\citep{Leibovich:2000ey}.  Practical application by the endpoint analyses 
is currently difficult because of integration of the rate 
in the $\Upsilon(4S)$ frame, which introduces significant smearing.  
With tagged $B$ samples, endpoint measurements can be made in the $B$ frame 
(with relaxed continuum suppression), which would allow for direct
 application of the weighted integral approach.  In principle, current $M_X$ analyses 
 can already use this approach, though experimental efficiency and
 $p_\ell$ cutoffs must be incorporated into the integrals.  All current analyses
 have, instead, relied on modeling an intermediate shape function.
 
The role of the shape function can be mitigated by restricting measurement
to regions of large $q^2$, where validity of the OPE expansion
is restored\citep{Bauer:2000xf,Bauer:2001rc}.   Restriction to the region kinematically 
forbidden to $\btoclnu$, $q^2>(M_B-M_D)^2$, however, 
introduces a low mass scale \citep{Neubert:2000ch,Neubert:2001ib}
into the OPE,  introducing uncertainties  of order $(\Lambda_{QCD}/m_c)^3$.
However, the combination of an $M_X$ with a looser $q^2$ requirement can suppress
$\btoclnu$ background yet still reduce shape function contributions.  Furthermore,
the $q^2$ restriction moves the parton level pole away from typical
$M_X$ cuts.   For a practical choice of region, $f(k_+)$ contributions are
suppressed but not negligible. The elimination of 
high energy hadronic final states by the $q^2$ restriction may 
exacerbate duality concerns.

A recent Belle analysis \citep{Kakuno:2003fk} in this region
employs a $p_\ell>1.2$ GeV$/c$ requirement followed
by an ``annealing'' procedure to sort reconstructed particles into  the ``signal'' 
and ``other'' $B$.   The analysis integrates over the rate in the region
 $M_X<1.7$ GeV and $q^2>8$ GeV$^2$ to extract $\Vub$, which again has the
 desired effect of minimizing dependence of the analysis on detailed modeling
 of the $\btoulnu$ process.   The signal to background ratio of the analysis,
about 1:6, does not approach that of the $B$ tag technique.
Belle finds a rate $\Delta{\cal B}$ in that $q^2$--$M_X$ region of
\begin{equation}
\frac{\Delta{\cal B}}{10^{-4}} = 7.37\pm0.89_{\mathrm stat} \pm 1.12_{\mathrm sys} \pm0.55_{c\ell\nu}
\pm0.24_{u\ell\nu}.
\end{equation}
I look forward to analysis of  this region with the significantly 
cleaner $B$ tag technique.

Extraction of $\Vub$ from these measurements has required modeling of $f(k_+)$ for
estimation of the fraction ($f$) of the rate in each region of
phase space.    The endpoint analyses use fractions estimated
by CLEO\citep{Bornheim:2002du}  based on an $f(k_+)$ derived from the CLEO 
$b\to s\gamma$ $E_\gamma$ spectrum.  CLEO employed several two--parameter functional forms
$f[\Lambda^{SF},\lambda_1^{SF}](k_+)$\citep{Bigi:1994it,Kagan:1998ym}
that were convolved with the parton-level $E_\gamma$ calculation in fits to the
measured spectrum between 1.5 and 2.8 GeV. These parameters 
are related to the HQET parameters of similar name, and play a similar role 
in evaluation of the rates.  At this time, however, we do not know the relationship
between the shape function parameters (or the moments of the
shape function) and the HQET nonperturbative parameters $\overline{\Lambda}=m_B-m_b$ and
$\lambda_1$\citep{neubertParams,Bauer:2003pi}.  The fact
that $\Lambda^{SF}$ and $\lambda_1^{SF}$ depend on the functional ansatz adopted,
while  the HQET parameters depend on the renormalization scheme, underscores
the current ambiguity.
 
\begin{figure}
 \includegraphics[width=0.9\columnwidth]{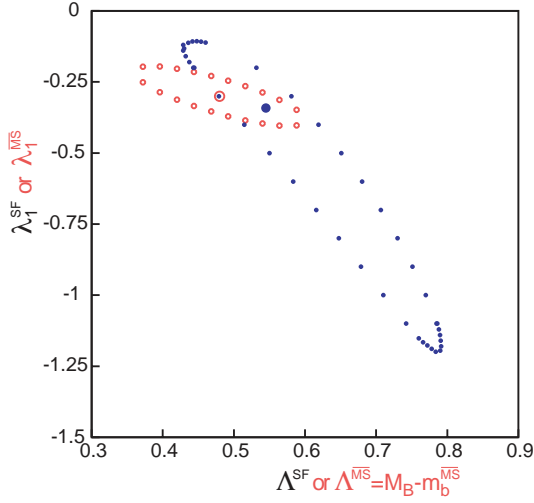}
  \caption{Shape function parameters from best fit to $b\to s\gamma$ photon energy
  spectrum and one standard deviation ``ellipse'' (solid blue circles).  Also shown is the
  ($\overline{\Lambda}^{\overline{MS}}$,$\lambda_1^{\overline{MS}}$) ellipse corresponding
  to the BaBar shape function parameter choice (open red circles). }
  \label{fig:bsgamma_ellipse}
\end{figure}

Figure~\ref{fig:bsgamma_ellipse} shows the best fit parameters\citep{bb:hennessy_private_com} 
and the one standard deviation contour for the exponential form\citep{Kagan:1998ym}.
The strong correlation between the two parameters
results from the interplay between the $b$ quark's effective mass (controlled by $\Lambda^{SF}$)
and kinetic energy (controlled by $\lambda_1^{SF}$) in determining
the {\em mean} of the $E_\gamma$ spectrum (and the mean energy available for the
final state in  $b\to u\ell\nu$).   The best fit corresponds to
$(\Lambda^{SF},\lambda_1^{SF})=(0.545,-0.342)$ and  the rate fractions $f_E=0.14$, $f_M=0.53$
and $f_{qM}=0.34$ for the CLEO $p_l>2.2$ GeV$/c$, the BaBar low $M_X$,
and the BELLE $M_X-q^2$ analyses, respectively.  The statistical uncertainty derives
from the extremes in the fractions found on the contour of 
Figure~\ref{fig:bsgamma_ellipse}.  The endpoint analysis was symmetrized by taking 
the average of those extremes, resulting in $f_E=0.13\pm0.02$, $f_M=0.52\pm0.12$ 
and  $f_{qM}=0.32\pm0.06$.  The fractions are almost completely correlated as one 
moves around the contour.  A small correction resulting from a re-optimization of background
normalization in the $b\to s\gamma$ analysis\citep{Chen:2001fj} (about 1/5 the assigned 
systematic) is applied. After including the remaining background subtraction systematic and smaller contributions from the $\alpha_s$ uncertainty and from variation of the $f(k_+)$ 
ansatz\citep{Bigi:1994it,Kagan:1998ym},  the rate fraction results are $f_E=0.13 \pm 0.03$ (with 
radiative corrections), $f_M=0.55\pm0.14$  and $f_{qM}=0.33\pm0.07$.  A more detailed
description is forthcoming\citep{gibbonsPRD}.

Two alternate approaches to shape function modeling  have been taken in experimental
studies. The BaBar $M_X$ analysis\citep{Aubert:2003zw} takes the $f(k_+)$ ansaetze 
just discussed,  but associates $(\Lambda^{SF},\lambda_1^{SF})$ with the HQET
parameters derived from spectral moments of the $b\to s\gamma$ and
$\btoclnu$ processes (see Figure~\ref {fig:bsgamma_ellipse}).  Note that while the
ellipse appears smaller, the $\overline{\Lambda}^{\overline{MS}}$--$\lambda_1^{\overline{MS}}$ 
correlation does not stabilize the average final state energy like the
$E_\gamma$-based correlation, resulting in $f_M$ uncertainties that are comparable to the $E_\gamma$ based uncertainties. The Belle $M_X$--$q^2$  analysis uses 
reference~\citep{Bauer:2001rc},  which employs a simpler form based 
on the  single parameter $\Lambda^{SF}/\lambda_1^{SF}$ estimated from
 the $m_b^{(1S)}$ mass and typical estimates for $\lambda_1$.  
These approaches rest on the assumption, now agreed to be 
incorrect\citep{neubertParams,Bauer:2003pi}, that an arbitrary renormalization
scheme can be used in the  parameterizations employed.
The associated uncertainty is difficult to assess, and has not
been included in the $\Vub$ determinations

Once the renormalization behavior of $f(k_+)$ is understood, the relationship between
HQET parameters, kinematic definitions of the $b$ quark mass, and moments of $f(k_+)$
will be better defined.  Moments of the $b\to c\ell\nu$ process and $m_b$ constraints
can then inform extraction of $f(k_+)$.  Correlations from the $m_b$ dependence
in the total rate and in the rate fractions must then be incorporated.  Given the complete
independence of the kinematic mass determination used for the total rate and
the effective quark mass in the $E_\gamma$--derived $f(k_+)$, coupled with
the large effective mass range sampled in the latter, a linear combination of those
two uncertainties\citep{Kakuno:2003fk} seems overly conservative for the present
$E_\gamma$--derived results.

Table~\ref{tab:inclusive_summary}, based on the Heavy Flavors Averaging 
Group (HFAG) summary, \citep{bb:hfag_vub}, summarizes
the full set of inclusive $\Vub$ results.  Given the strong correlations 
among the rate fractions, $\Vub$ comparisons are meaningful only for 
results evaluated with common theoretical input.
The $E_\gamma$--derived shape function is currently the best motivated
theoretically and has the most complete categorization of uncertainties.  I therefore
adjust recent results and $f(k_+)$--related uncertainties to use the rate fractions discussed above.
The errors do not include uncertainties from potentially large theoretical corrections that have been categorized but not calculated, as discussed below.

\ifthenelse{\boolean{beauty}}{
\begin{table}
\begin{tabular}{lll} \hline
& $\Vub (10^{-3})$ & \\ \hline
ALEPH\citep{Barate:1998vv}  & $4.12 \pm 0.67 \pm 0.76$ & neur. net \\
L3\citep{Acciarri:1998if}           & $5.70 \pm 1.00 \pm 1.40$ & cut, count \\
DELPHI & $4.07 \pm 0.65 \pm 0.61$ & $M_X$  \\
OPAL\citep{Abbiendi:2001qx}     & $4.00 \pm 0.71 \pm 0.71$ & neur. net \\
LEP Avg.\citep{hfworkinggroup} &  $4.09 \pm 0.37 \pm 0.56$ & \\
CLEO\citep{Bornheim2002triplediff}     & $4.05 \pm 0.61 \pm 0.65$ & triple diff. \\ 
BELLE   & $5.00 \pm 0.64 \pm 0.53$ &  $M_{X}$ \\ \hline 

CLEO     & $4.11 \pm 0.34 \pm 0.46 \pm 0.28$ &  $p_\ell> 2.2$\\
BaBar  & $4.31 \pm 0.28 \pm 0.49 \pm 0.30$ & $p_\ell> 2.3$ \\
BELLE   & $3.99 \pm 0.23 \pm 0.45 \pm 0.27$ & $p_\ell> 2.3$ \\
BELLE   & $4.63 \pm 0.48 \pm 0.48 \pm 0.32$ & $^{M_X<1.7}_{q^2>8}$ \\
BaBar  & $4.79 \pm 0.40 \pm 0.60 \pm 0.33$ & $M_{X}<1.55$

\end{tabular}
\caption{Summary of inclusive $\Vub$ results.  The errors on the first group
are experimental and theoretical uncertainties, respectively. The second group,
adjusted to a common $f(k_+)$, lists the total experimental,  $f(k_+)$--related, 
and $\Gamma_{\mathrm tot}$ uncertainties. The 
two groups are not directly comparable.}
\label{tab:inclusive_summary}
\end{table}
}{
\begin{table}
\begin{tabular}{lcl} \hline
& $\Vub (10^{-3})$ & \\ \hline
ALEPH  & $4.12 \pm 0.67 \pm 0.76$ & neural net \\
L3           & $5.70 \pm 1.00 \pm 1.40$ & cut and count \\
DELPHI & $4.07 \pm 0.65 \pm 0.61$ & $M_X$  \\
OPAL     & $4.00 \pm 0.71 \pm 0.71$ & neural net \\
LEP Average &  $4.09 \pm 0.37 \pm 0.56$ & \\
CLEO     & $4.05 \pm 0.61 \pm 0.65$ & $d\Gamma/dq^2dM_X^2dE_\ell$ \\
BELLE   & $5.00 \pm 0.64 \pm 0.53$ &  $M_{X}$, $D^{(*)}\ell\nu$ tag \\
CLEO     & $4.11 \pm 0.13 \pm 0.31 \pm 0.46 \pm 0.28$ &  2.2 < p < 2.6 \\
BABAR  & $4.31 \pm 0.20 \pm 0.20 \pm 0.49 \pm 0.30$ & 2.3 < p < 2.6  \\
BELLE   & $3.99 \pm 0.17 \pm 0.16 \pm 0.45 \pm 0.27$ & 2.3 < p < 2.6  \\
BELLE   & $4.63 \pm 0.28 \pm 0.39 \pm 0.48 \pm 0.32$ & $M_X<1.7$, $q^2>8$ \\
BABAR  & $4.79 \pm 0.29 \pm 0.28 \pm 0.60 \pm 0.33$ & $M_{X}<1.55$
\end{tabular}
\caption{Summary of all inclusive $\Vub$ measurements.  The last five measurements
are incorporated into the combination presented here.  The errors on the first seven
values are XXX and XXX, respectively (what are these HFAG?!)  The errors on the last
five are statistical, experimental systematic, shape function uncertainty , and the $\Gamma_{\mathrm tot}$ uncertainties (OPE, $m_b$, duality) added in quadrature. Note
that the total uncertainties in the first and second groups are {\em not} directly comparable
as they have not been evaluated with identical theoretical inputs.}
\label{tab:inclusive_summary}
\end{table}
}

\subsection{Combining inclusive information}

Evaluation of the total uncertainty on $\Vub$ remains problematic because of
three main theoretical complications (\citep[see, {\it e.g.},][]{Luke:2003nu,Ligeti:2003hp}.  
The
first arises from subleading (higher twist) contributions to the OPE
resummation\citep{Leibovich:2002ys,Bauer:2002yu,Neubert:2002yx,Bauer:2001mh},
which are not universal.  Hence with
use of $b\to s\gamma$ to constrain $f(k_+)$, there exist subleading contributions
both to the use of a shape function in $\btoulnu$ itself and 
to the derivation of $f(k_+)$ from $b\to s\gamma$.  The subleading 
contributions, formally of order  $\Lambda_{QCD}/m_b$, can be as large as $\sim15\%$.  
Indeed, a partial estimate\citep{Neubert:2002yx} for the endpoint region finds corrections 
that are approximately  the same size as the other combined uncertainties.

The second contribution, from ``weak annihilation''  processes\citep{Bigi:1994bh,Voloshin:2001xi}, 
is formally of order  $(\Lambda_{QCD}/m_b)^3$ but receives a
$16\pi^2$ enhancement.
The contribution, which requires factorization violation to be nonzero, is expected to be 
localized near $q^2\sim m_b^2$. This
results in further enhancement of the effect on $\Vub$ measurements.  For
the endpoint region, an effect on the total rate of 2-3\% (for factorization
 violation of about 10\%), corresponds to 20-30\% on the endpoint rate. 

Finally, while global quark hadron duality is well-motivated for spectral moments, 
the OPE cannot predict the detailed inclusive spectra.  The extent of
violation of quark--hadron duality locally
depends on the size and nature of the region of phase space considered: 
including a large fraction of the rate is best.
The associated uncertainty is difficult to assess.

The problems outlined present a considerable obstacle to a meaningful
average of the inclusive results.   Results with a potentially large
bias will be included with neither correction nor meaningful uncertainty:  $\Vub$ will be 
biased and have an unreliable uncertainty.
As an alternative, we can choose a region of phase space
that provides a reasonable compromise among the unknown contributions.
The choice is inherently subjective given the different viewpoints
within the theory community (\citep[see, {\it e.g.},][]{Bigi:2002qq,Bauer:2001rc}).  In this reviewer's
opinion, the opportunity to bound experimentally the uncertainties we know of,
thereby providing as complete an uncertainty estimate as possible,
is more important than achieving the smallest statistical precision.  Currently,
I find the region restricted to low $M_X$ and higher $q^2$ the most compelling. It has
reduced corrections from $f(k_+)$ and hence from subleading contributions, 
yet has a sufficient fraction of the spectrum to dilute weak annihilation and local quark hadron duality concerns.  The low $M_X$ and $p_l$ endpoint regions, in which one or more of the
corrections is more pronounced, then play critical
roles in limiting the uncertainty in this region.  I stress that estimating a complete inclusive
uncertainty is of fundamental importance, and hence that I consider measurements in
all three regions of equal importance.  Indeed, I view a single coherent analysis of all three
phase space regions simultaneously an important milestone for both $B$ factories, 
particularly with application of the powerful and clean $B$ tag samples.  

For now, however, only BELLE has contributed a result for this region of phase space, 
and I quote that result as my  ``central value'':
\begin{eqnarray}
\lefteqn{\Vub/10^{-3} = 4.63 \pm 0.28_{\mathrm stat} \pm 0.39_{\mathrm sys} \pm } \\ \nonumber 
& & 0.48_{f_{qM}} \pm 0.32_{\Gamma_{\mathrm thy}} 
\pm \sigma_{\mathrm WA} \pm \sigma_{\mathrm SSF} \pm \sigma_{\mathrm LQD}.
 \end{eqnarray}
New measurements  in this region can be easily combined
when available, and will improve the experimental uncertainties.
We must determine the  uncertainties for weak annihilation (WA), subleading
shape function corrections (SSF) and local quark hadron duality (LQD) within this region.
Note that the data are not yet precise enough to draw conclusions regarding
the presence or absence of these corrections; we use them to provide bounds.

Each phase space region considered should largely contain the WA contribution, 
which  will be most (least) diluted in the low $M_X$  (endpoint) region.  For a 
neglected WA contribution, comparison of $\Vub$ from these two regions would 
predict a bias in the $M_X$, $q^2$ region to be
\begin{equation}
[(1-f_{qM})/f_{qM}][f_e f_M/(f_M-f_e)] \approx 0.39
\label{eq:wa_scale}
\end{equation}
of the observed difference. The quoted value, which is model dependent,  
is based on the fractions found  for the $E_\gamma$--derived $f(k_+)$.
Comparison of the CLEO endpoint and BaBar low $M_X$ values, taking into consideration
the almost total correlation in the shape function and $\Gamma_{tot}$ uncertainties,
yields $\Delta|V_{ub}|/10^{-3} = 0.69 \pm 0.53$.
I take the larger of the central value and error and scale 
according to Eq.~\ref{eq:wa_scale} to obtain $\sigma_{WA} \approx 0.27$.

To estimate $\sigma_{SSF}$, I assume that subleading corrections scale like 
the fractional change in the rate prediction ($\Delta \Gamma/\Gamma$) 
that  $f(k_+)$ induces  relative to the parton-level calculation.   Comparison of the low 
$M_X$ region ($[\Delta\Gamma/\Gamma]_M\sim 0.15$) to the combined $M_X$, $q^2$
region ($[\Delta \Gamma/\Gamma]_{qM}\sim-0.075$) with theory correlations considered,
gives $\Delta|V_{ub}|/10^{-3} = 0.16 \pm 0.63$.   Scaling  by $|(\Delta \Gamma/\Gamma)_{qM}/(\Delta \Gamma/\Gamma)_M|=0.49$, which is model dependent,
we obtain $\sigma_{\mathrm SSF} \approx 0.31$.

Finally, to bound the local duality uncertainty, I assume
that the error scales with the rate fraction $f$ as  $(1-f)/f$ 
({\em ad hoc}, but goes to zero for full phase space and diverges for use of
the detailed spectra).  To obtain an estimate, I compare the CLEO $p_\ell>2.2$ GeV/$c$ region
 to the average of BaBar and BELLE in the $p_\ell>2.3$ GeV/$c$ region
and apply the subleading correction
estimates\citep{Neubert:2002yx}  ($+0.27\times10^{-3}$) to minimize potential cancelation between
duality violation and subleading corrections.  This yields $(\Vub^{2.3}-\Vub^{2.2}+0.27)/10^{-3}=0.29\pm0.38$.  Scaling the uncertainty by
\begin{equation}
\frac{(1-f_{qM})/f_{qM}}{(1-f_{2.3})/f_{2.3}-(1-f_{2.2})/f_{2.2}} \approx 0.29
\end{equation}
based on the fractions in Table~\ref{tab:endpoint_rates} 
gives $\sigma_{\mathrm LQD} \sim 0.11$.

From this combination of information, we thus find
\begin{eqnarray}
\lefteqn{\Vub/10^{-3}  =  4.63 \pm 0.28_{\mathrm stat} \pm 0.39_{\mathrm sys} \pm} \\ \nonumber 
 & &  0.48_{f_{qM}} \pm 0.32_{\Gamma_{\mathrm thy}} 
 \pm 0.27_{\mathrm WA}  \pm 0.31_{\mathrm SSF}  \pm 0.11_{\mathrm LQD}
 \end{eqnarray}
for a total theory error of 15\%.  Since experimental uncertainties dominate, quadrature
addition seems reasonable.   The limits presented
here can be improved in robustness ( {\it e.g.} considering
potential cancellations among effects), through more sophisticated scaling estimates,
and  through improved and additional measurements.  Improvement of the $b\to s\gamma$ photon
energy spectrum is key.  Measurements of $D^0$ versus $D_s$ semileptonic widths 
and comparison of neutral and charged $B$ decay can help limit WA 
contributions\citep{Voloshin:2001xi}.   Finally, improved theory
for the scaling of the effects over phase space could allow development of a 
procedure for simultaneous extraction of $\Vub$ and the corrections, with all experimental information 
contributing  directly to $\Vub$, or could shift the choice of ``preferred'' region.

\subsection{Exclusive measurements of $\btoulnu$}

I devote considerably less time to discussion of exclusive determination
of $\Vub$ -- not because it is less important but because the story
is simpler.   Theoretical issues center on determination of the
form factors (FF) involved in the decays.  For $B\to\pi\ell\nu$, for example,
one has\citep{Gilman:1990uy}
\begin{equation}
\frac{d\Gamma(B\to\pi\ell\nu)}{dq^2\,d\cos\theta_\ell} =
\Vub^2\frac{G_F^2 p_\pi^3}{32\pi^3}\sin^2\theta_\ell|f_+(q^2)|^2
\end{equation}
with only the single form factor $f_+(q^2)$ for massless leptons.  Final states with
a vector meson depend on three form factors. The measured rates depend on the 
variation of the form factors with $q^2$ (``shape'') and their relative 
normalizations\citep{Gibbons:2003gq}, 
while extraction of $\Vub$ depends both on their shape and absolute normalization.

A large variety of calculations exist.  For extraction of $\Vub$, the focus has sharpened
onto the QCD--based calculations of lattice QCD  (LQCD)\citep{Abada:1994dh,Allton:1995ui,DelDebbio:1998kr,Hashimoto:1998sr,Ryan:1998tj,Ryan:1999kx,Lellouch:1999dz,Bowler:1999xn,Becirevic:1999kt,Aoki:2000by,Abada:2000ty,El-Khadra:2001rv,Aoki:2001rd} and
light cone sum rules (LCSR)\citep{Ball:1997rj,Ball:1998kk,Khodjamirian:1997ub,Khodjamirian:2000ds,Bakulev:2000fb,Huang:2000hs,Wang:2001mi,Wang:2001bh,Ball:2001fp}.  Only
quenched LQCD FF calculations are available for $b\to u\ell\nu$, and these have sensitivity
only in the range $q^2\gtrsim 16$ GeV$^2$.  Approximations made in the LCSR calculations
are only valid for $q^2\lesssim 16$ GeV$^2$.  A recent summary\citep{Battaglia:2003in} shows
reasonable agreement where comparison is possible.   The FF's have
also been evaluated using many quark-- or parton--model based 
techniques\citep{Wirbel:1985ji,Korner:1988kd,Isgur:1989gb,Scora:1995ty,Melikhov:1996xz,Beyer:1998ka,Faustov:1996bf,Demchuk:1997uz,Grach:1996nz,Riazuddin:2000ae,Melikhov:2000yu,Feldmann:1999sm,Flynn:2000gd,Beneke:2000wa,Choi:1999nu}.  Uncertainties
in these models are difficult to assess, and they exhibit a broad
variation in shape.  Finally, various studies of FF's based on constraints from dispersion relations,
unitarity or Heavy Quark Symmetry have been 
made\citep{Kurimoto:2001zj,Ligeti:1996yz,Aitala:1998cm,Burdman:1997kr,Lellouch:1996yv,Mannel:1998kp}.

Recent work has helped assessment of the reliability of the FF's based on the
available LCSR  and quenched  (no light quark loops in the propagators) LQCD calculations.
Analysis\citep{Lange:2003jz,Ball:2003bf,Beneke:2003pa,Lange:2003pk} of the
LCSR approach within the framework of soft collinear effective theory (SCET) 
has sparked debate regarding potential contributions missing from LCSR.
The $B\to\rho\ell\nu$ FF's, in particular, may be overestimated in LCSR,
biasing $\Vub$ low.  Unquenched LQCD calculations have begun to appear,
and comparison to experiment shows much better agreement with data (few percent)
than for quenched results\citep{Davies:2003ik}.  While work has begun on unquenched
FF's, initial results are limited to valence quark masses $\sim m_s$.
Initial results\citep{Bernard:2003gu,Okamoto:2003ur} are compatible with
the $\sim15\%$  uncertainties assigned for  the quenching approximation.

To date, all exclusive measurements employ detector hermeticity to
estimate $p_\nu$, which allows full reconstruction of the decays.
Two general strategies have been taken.  The CLEO\citep{Behrens:1999vv}
and BaBar\citep{Aubert:2003zd} $B \to \rho \ell \nu$ analyses emphasize
higher efficiency and employ relatively loose event cleanliness and
$\nu$-consistency criteria.  With the resulting background levels, these
analyses are primarily sensitive in the region $p_\ell>2.3$ GeV$/c$. 
To extract total branching fractions and $\Vub$, these analyses survey a 
variety of FF models, including LQCD and LCSR calculations 
extrapolated over the full $q^2$ range.

\begin{figure}[b]
 \includegraphics[width=0.8\columnwidth]{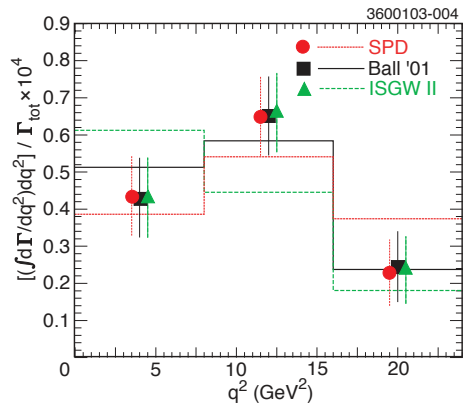}
 \source{\citet{Athar:2003yg}}
  \caption{The CLEO $B^0\to \pi^-\ell^+\nu$ partial branching fractions based
  on three disparate FF models.}
  \label{fig:dGdq2}
\end{figure}

The simultaneous $B \to \pi \ell  \nu$ and $B \to \rho \ell  \nu$ measurement
by CLEO \citep{Athar:2003yg}, on the other hand, applies strict criteria to achieve
acceptable background levels over a broad $p_\ell$ range.  With sensitivity
down to 1.0 GeV$/c$ (1.5 GeV$/c$) for $\pi \ell  \nu$ ($\rho \ell  \nu$), this analysis
was able to extract independent rates in three $q^2$ bins.   This
eliminated model dependence of the  measured $\pi \ell  \nu$ rates 
(Figure~\ref{fig:dGdq2}) and halved
that for $\rho\ell\nu$ rates (see Ref.~\citep{Gibbons:2003gq} for a discussion
of model dependence).  Furthermore, the analysis permits extraction of
$\Vub$ from the LQCD and LCSR FF's within their valid $q^2$ ranges and so without
additional modeling.

Table~\ref{tab:exclusiveVub} summarized the results for $\Vub$ from exclusive 
measurements.  Averages of the CLEO results are given with and without 
the low $q^2$ region for $\rho\ell\nu$ (for which LCSR validity is under debate).
Note that FF-related uncertainties have been treated as completely correlated
in the LCSR and LQCD averages to remain conservative.  The LQCD uncertainties
for the $\pi$ ($\rho$) modes include 15\% (20\%) quenching uncertainties,
which are called out separately in the LQCD references.
Ref.~\citep{Gibbons:2003gq} presents a more complete review of recent $B\to X_u\ell\nu$ 
branching fractions,  including measurements from which $\Vub$ has not yet been extracted.
Of note is evidence for $B^+\to\omega\ell^+\nu$\citep{Abe:2003rh} and
$B^+\to\eta\ell^+\nu$\citep{Athar:2003yg}. 

\ifthenelse{\boolean{beauty}}{
\begin{table}
\begin{tabular}{llcl} \hline
 & $\Vub (10^{-3})$ & $q^2$ range & FF \\ \hline
 & & (GeV$^2$) & \\
CLEO '00 $\rho$ & $3.23^{+0.33}_{-0.35}\pm0.58$ & all & survey \\
BaBar '01 $\rho$ & $3.64 \pm0.33^{+0.39}_{-0.56}$ & all & survey \\
CLEO '03 $\pi$ & $3.33 \pm0.28^{+0.57}_{-0.40}$ & $<16$  & LCSR \\ 
CLEO '03 $\pi$ & $2.88 \pm0.63^{+0.48}_{-0.39}$ & $>16$  & LQCD \\
CLEO '03 $\pi$ & $3.24 \pm0.26^{+0.56}_{-0.40}$ & \multicolumn{2}{l}{average} \\
CLEO '03 $\rho$ & ${2.67^{+0.47}_{-0.50}}^{+0.50}_{-0.39}$ & $<16$  & LCSR \\ 
CLEO '03 $\rho$ & ${3.34^{+0.42}_{-0.48}}^{+0.69}_{-0.62}$ & $>16$  & LQCD \\
CLEO '03 $\rho$ & ${3.00^{+0.36}_{-0.41}}^{+0.56}_{-0.47}$ & \multicolumn{2}{l}{average} \\
CLEO '03 $\pi$,$\rho$ & ${3.17^{+0.23}_{-0.24}}^{+0.53}_{-0.39}$ & \multicolumn{2}{l}{average} \\
CLEO '03 $\pi$,$\rho$ & $3.26 \pm0.24^{+0.54}_{-0.39}$ & \multicolumn{2}{l}{average, no $\rho$ LCSR}
\end{tabular}
\caption{Summary of exclusive $\Vub$ measurements.  
The errors listed are the statistical and experimental systematic uncertaines combined
in quadrature,  and all form factor uncertainties, respectively.
In the CLEO '03 averages, the LQCD and LCSR uncertainties have been treated as correlated.}
\label{tab:exclusiveVub}
\end{table}
}{
\begin{table}
\begin{tabular}{llccl} \hline
& mode & $\Vub (10^{-3})$ & $q^2$ range & FF \\ \hline
CLEO '00 $\rho$ & $3.23\pm0.24^{+0.23}_{-0.26}\pm0.58$ & all & model survey \\
BaBar '01 $\rho\$ & $3.64\pm0.22\pm 0.25^{+0.39}_{-0.56}$ & all & model survey \\
CLEO '03 $\pi$ & $3.33\pm 0.24 \pm0.15\pm 0.06 \;^{+0.57}_{-0.40}$ & $q^2<16$ GeV$^2$ & LCSR \\ 
CLEO '03 $\pi$ & $2.88\pm 0.55\pm 0.30\pm 0.18\;^{+0.45}_{-0.35}$ & $q^2>16$ GeV$^2$ & LQCD \\
CLEO '03 $\pi$ & $3.24\pm0.22\pm 0.13\pm0.09 ^{+0.55}_{-0.39}$ & \multicolumn{2}{l}{average} \\
CLEO '03 $\rho$ & $2.67\pm 0.27\;^{+0.38}_{-0.42} \pm 0.17 \;^{+0.47}_{-0.35}$ & $q^2<16$ GeV$^2$ & LCSR \\ 
CLEO '03 $\rho$ & $3.34\pm 0.32\;^{+0.27}_{-0.36}  \pm 0.47 \;^{+0.50}_{-0.40}$ & $q^2>16$ GeV$^2$ & LQCD \\
CLEO '03 $\rho$ & $3.00\pm0.21^{+0.29}_{-0.35}\pm0.28 ^{+0.49}_{-0.38}$ & \multicolumn{2}{l}{average} \\
CLEO '03 $\pi+\rho$ & $3.17\pm0.17^{+0.16}_{-0.17}\pm0.03 ^{+0.53}_{-0.39}$ & \multicolumn{2}{l}{average} \\
CLEO '03 $\pi+\rho$ & $3.26\pm0.19\pm0.15\pm0.04 ^{+0.54}_{-0.39}$ & \multicolumn{2}{l}{average, no $\rho$ LCSR}
\end{tabular}
\caption{Summary of exclusive $\Vub$ measurements.  
For the CLEO '00 and BaBar '01 measurements, the errors arise from
 statistical, experimental systematic  and form factor modeling uncertainties, respectively.
For the CLEO '03 measurements, the errors arise from statistical, experimental systematic  $\rho\ell\nu$ form factor, and LQCD and LCSR calculation uncertainties, respectively.
In the CLEO '03 averages, the LQCD and LCSR uncertainties have been treated as correlated.}
\label{tab:exclusiveVub}
\end{table}
}

Potential exists for significant correlation among the dominant experimental 
systematics\citep{Gibbons:2003gq}, so the results have been averaged assuming full
correlation.   The earlier results\citep{Behrens:1999vv,Aubert:2003zd},
 which depend more heavily on modeling, are deweighted by 5\%. The average
 yields
\begin{equation}
|V_{ub}| = (3.27 \pm 0.13 \pm0.19 ^{+0.51}_{-0.45}) \times 10^{-3}
\end{equation}
where
the  errors arise from statistical, experimental systematic  and form factor
uncertainties, respectively.  
Should the LCSR form factors prove to be overestimated, I also 
average without including information using $q^2<16$ GeV$^2$ in
$\rho\ell\nu$, yielding $|V_{ub}| = (3.26\pm0.19\pm0.15\pm0.04 ^{+0.54}_{-0.39}) \times 10^{-3}$,
where the  errors are statistical, experimental systematic  $\rho\ell\nu$ form factor
uncertainties, and LQCD and LCSR uncertainties. 

The future for exclusive determinations of $\Vub$ appears promising. 
The large $B$ tag samples being collected should allow
significant improvement in  $\nu$ resolution and reduction of 
backgrounds and experimental systematics.  Unquenched LQCD 
calculations are underway and will eliminate the primary, poorly controlled, source of uncertainty.    Recent advances may also allow 
use of the full $q^2$ range for  extraction of $\Vub$ with LQCD\citep{Foley:2002qv,Boyle:2003ui}.   For both LQCD and experiment, $\pi\ell\nu$ appears to be the golden mode --
even the $B^*$ pole now appears manageable\citep{El-Khadra:2001rv}.
$B\to \eta\ell\nu$ will provide a valuable cross-check.
The $\rho\ell\nu$ mode will be more problematic for high precision. The broad $\rho$ width leaves
experiments open to larger backgrounds, including poorly-understood
nonresonant $\pi\pi$ contributions.  In unquenched LQCD calculations the $\rho$ is unstable, and
methods for accommodating the high energy $\pi\pi$ final state have yet to be developed.  
The $\omega\ell\nu$ mode may prove more
tractable. Agreement between accurate $\Vub$  determinations from $\pi\ell\nu$, $\eta\ell\nu$ 
and $\omega\ell\nu$ will add confidence overall. 

\subsection{Inclusive, exclusive averaging}
Until recently, I have argued against averaging of inclusive and
exclusive results because of outstanding uncertainties in the former and no
checks on the latter's theory.  With the experimental bounds on uncertainties
determined above, and some clarification of the reliability of LCSR and of
the quenching uncertainties in LQCD, my major concerns are being addressed.
I therefore combine the inclusive and exclusive results and find
\begin{equation}
\Vub = (3.67\pm0.47)\times 10^{-3}.
\end{equation}
Excluding results based on data below $q^2<16$ GeV$^2$ in the $\rho\ell\nu$
yields a similar result: $\Vub = (3.70\pm0.49)\times 10^{-3}$.  Inclusive and
exclusive averaging will likely remain controversial in the short term.  However,
with the progress expected both inclusive and exclusive measurements and 
theory, I anticipate a noncontroversial value of $\Vub$, with an uncertainty
bettering the 13\% presented here, in the next few years.


\begin{theacknowledgments}
I would like to thank A.~Kronfeld, B.~Lange, G.P.~Lepage, Z.~Ligeti, M.~Luke and M.~Neubert
for their input regarding the theory of $\Vub$ determinations.  Special thanks to D.~Cronin-Hennessy and E.~Thorndike for their analysis of the rate fractions derived from the $E_\gamma$ spectrum.
 \end{theacknowledgments}


\bibliographystyle{aipproc}   

\bibliography{VubReferences}

\IfFileExists{\jobname.bbl}{}
 {\typeout{}
  \typeout{******************************************}
  \typeout{** Please run "bibtex \jobname" to optain}
  \typeout{** the bibliography and then re-run LaTeX}
  \typeout{** twice to fix the references!}
  \typeout{******************************************}
  \typeout{}
 }

\end{document}